\documentclass[twocolumn,amsmath,amssymb,showpacs]{revtex4}
\usepackage{graphicx}

\begin{document}

\title{Resolving different pairing states in Weyl superconductors through the single-particle spectrum}

\author{Tao Zhou$^{1,2,3}$}
\email{tzhou@m.scnu.edu.cn}

\author{Yi Gao$^{4}$}

\author{Z. D. Wang$^{3}$}
\email{zwang@hku.hk}

\affiliation{
$^{1}$Guangdong Provincial Key Laboratory of Quantum Engineering and Quantum Materials,
and School of Physics and Telecommunication Engineering,
South China Normal University, Guangzhou 510006, China\\
$^{2}$College of Science, Nanjing University of Aeronautics and Astronautics, Nanjing 210016, China.\\
$^{3}$Department of Physics and Center of Theoretical and Computational Physics, The University of Hong Kong, Pokfulam Road, Hong Kong, China.\\
$^{4}$Department of Physics and Institute of Theoretical Physics,
Nanjing Normal University, Nanjing, 210023, China.
}

\date{\today}
\begin{abstract}

We study theoretically single-particle spectra of Weyl superconductors. Three different superconducting pairing states are addressed, which are the BCS-type states with the $s$-wave pairing symmetry and the $p+ip$-wave pairing symmetry, and the FFLO pairing state. We elaborate that these three states can be resolved well based on the bulk and surface spectral functions as well as the local density of states. The single impurity effect is also explored, which may help us to differentiate the BCS-type pairing states and the FFLO state further.

\end{abstract}
\maketitle

\section{introduction}

The superconductivity in Weyl semimetal systems has been paid an intensive attention. Experimentally, the pressure induced superconductivity
was reported in the materials of WTe$_2$~\cite{pan,kang,chan} and WoTe$_2$~\cite{yqi}.
The surface superconductivity without the pressure effect was detected by the point contact
spectroscopy in MoTe$_2$ material~\cite{nai}.
The above parent materials have been predicted to be Weyl semimetals~\cite{sol,ysun}.
Recently, the (Ta,Nb)(As,P) family had been identified to be Weyl semimetals~\cite{weng,huang,syxu,bqlv,bqlv1,syxu1,syxu2}.
The signature of superconductivity was revealed by the point contact tunneling
spectrum on the TaAs material~\cite{hwang}. The intrinsic superconductivity was realized in the TaP material under pressure~\cite{yuli}.
It was reported that superconductivity can be induced to the NbAs material through ion irradiation~\cite{bachm}.
Theoretically, it was proposed that the
Nb-doped Bi$_2$Se$_3$ and the heavy fermion superconductor UPt$_3$ are Weyl superconductors~\cite{yuan,yana}.
Apart from the natural materials, it was also proposed that superconductivity may be realized through doping, the proximity effect, or applying a magnetic field. Physical properties of Weyl superconductors and their potential applications have attracted broad interest in the past few years~\cite{meng,cho,wei,bed,tzhou,das,chen,kha,yang,liu,blu,jian,kim,baru1,lhao,yli,jfa,taik,baru}.

A minimum model for describing the doped Weyl semimetals generates two
pockets around the two Weyl nodes~\cite{cho,wei,bed,tzhou}. Two competing pairing states were before proposed when the superconductivity is induced.
One is the inter-pocket pairing state where the momentum of the Cooper pair is zero, known as the BCS pairing state~\cite{wei,bed,tzhou}. The other is the intra-pocket pairing state with finite momentum Cooper pairs, named as the Fulde-Ferrell-Larkin-Ovchinnikov (FFLO) state~\cite{cho}. It was proposed that the FFLO state should win over the BCS state, in view of that the superconducting pairing is within a thin shell around the Fermi surface~\cite{cho}. While the BCS pairing was proposed to be favored when the pairing in the whole Brillouin zone was considered~\cite{tzhou}.
Note that the above two proposed pairing states are both in even channel. An odd channel (triplet pairing) BCS pairing state was also proposed~\cite{wei,bed}.
It was also proposed in Ref.~\cite{bed} that the FFLO state might win over the odd-channel BCS state when both inversion and time-reversal
symmetry are broken. Therefore, no consensus about the ground state of the Weyl superconductor has so far been reached.

The superconductivity in Weyl systems may be related to the realization of Majorana bound states, which has attracted considerable interest
due to their potential applications in topological quantum computation~\cite{naya}. Besides, the realization of the FFLO state is also an important issue in the studies of superconductivity. The FFLO state was proposed about fifty years ago~\cite{ful,lar}, while its existence has not been confirmed yet despite intensive efforts.
Especially, it was indicated theoretically that the space-time supersymmetry emerges when the Weyl system transits to the FFLO state, which is also of fundamental interest~\cite{sjian}.
Therefore, to determine the ground state of Weyl superconductors is rather important, while, theoretically, the results may depend on the parameters, the model, the approximation used, and the pairing mechanism. Therefore, identifying the most favorable pairing state of a Weyl superconductor directly  appears to be challenging. More detailed information is needed to resolve the different pairing states.

The single-particle spectrum has been used widely for investigating the electronic structure and determining the pairing states of unconventional superconductors. Experimentally, it can be measured by the angle-resolved photoemission spectroscopy (ARPES)~\cite{dama} or the scanning tunneling microscopy (STM)~\cite{oys}. Theoretically, it is obtained by calculating the imaginary part of the single-particle Green's function. Therefore, an effective link for theoretical calculations
and experimental observations may be established through analyzing the single-particle spectrum. For Weyl superconductors,
systematic investigations about its single-particle spectrum are still awaited.
Thus it is timely and useful to look into this issue theoretically, which enable us to understand the electronic structure and topological features of this family. Moreover, it may help to resolve different pairing states when superconductivity is present.

In this paper, motivated by the above considerations, we have studied theoretically the single-particle spectrum of Weyl superconductors. Three different pairing states, i.e., the $s$-wave BCS pairing state~\cite{tzhou}, the $p+ip$ BCS pairing state~\cite{wei}, and the FFLO pairing state~\cite{cho}, have been considered. The spectral function in the momentum space and the local density of states (LDOS) are both explored and may be used to distinguish different pairing states. We also study the LDOS near a point nonmagnetic impurity and propose that the in-gap impurity states may be used to resolve the pairing states further.

The rest of the paper is organized as follows.
In Sec. II, we introduce
the model and present the relevant formalism. In Sec. III, we
report numerical calculations and discuss the obtained
results. Finally, we give a brief summary in Sec. IV.

\section{Model and formalism}

We start from a lattice model in the real space to describe the Weyl superconductor, consisting of both the Weyl metal and the superconducting pairing terms,
\begin{equation}
H=H_{W}+H_{sc}.
\end{equation}
  Here $H_{W}$ is an effective lattice model for the Weyl metal~\cite{tzhou}, expressed as,
\begin{eqnarray}
H_W=&-\sum\limits_{{\bf i}}\sum\limits_{\alpha}[C^{\dagger}_{\bf i}t_\alpha \sigma_3  C_{{\bf i+\hat{\alpha}}}+C^{\dagger}_{\bf i} (h\sigma_3-\mu\sigma_0)C_{{\bf i}}+h.c.]\nonumber\\
&+\sum_{\bf i}\lambda(C^\dagger_{\bf i}\sigma_1 C_{{\bf i}+\hat{x}}+ C^{\dagger}_{{\bf i}}\sigma_2 C_{{\bf i}+\hat{y}}+h.c.),
\end{eqnarray}
with $C_{\bf i}=(c_{{\bf i}\uparrow},c_{{\bf i}\downarrow})^{T}$. $\sigma_0$ and the $\sigma_{1-3}$ are the identity matrix and Pauli matrix, respectively. ${\bf i}=(x,y,z)$ represents a site on the three dimensional cubic lattice. $\alpha=$$\hat{x}$, $\hat{y}$, and $\hat{z}$, are
the base vectors along $x$, $y$, and $z$ directions, respectively.
While $H_{SC}$ represents the superconducting pairing term. Here the considered pairing states include the $s$-wave BCS state, the $p+ip$-wave BCS state with equal spin, and the FFLO state.

The $s$-wave pairing state is expressed as,
\begin{equation}
H^{s}_{SC}=\sum_{\bf i}(\Delta_0c^{\dagger}_{{\bf i}\uparrow}c^{\dagger}_{{\bf i}\downarrow}+h.c.).
\end{equation}

The $p+ip$-wave pairing state is expressed as,
\begin{equation}
H^{p+ip}_{SC}=\sum_{{\bf ij}\sigma}(\Delta_{\bf ij}c^{\dagger}_{{\bf i}\sigma}c^{\dagger}_{{\bf j}\sigma}+h.c.),
\end{equation}
where ${\bf j}={\bf i}\pm \hat{y}(\hat{x}) $. $\Delta_{\bf ij}=\pm\frac{\Delta_0}{2}$ and $\pm \frac{i\Delta_0}{2}$ for the cases of ${\bf j}={\bf i}\pm\hat{y}$ and ${\bf j}={\bf i}\pm\hat{x}$, respectively.

The FFLO pairing state is expressed as,
\begin{equation}
H^{FFLO}_{SC}=\sum_{\bf i}[\Delta_0\cos({\bf R_i}\cdot {\bf Q_f})c^{\dagger}_{{\bf i}\uparrow}c^{\dagger}_{{\bf i}\downarrow}+h.c.].
\end{equation}
${\bf Q_f}$ is the net momentum of a Cooper pair. In the FFLO state, the pairing order parameters vary periodically in the real space and the period
$r$ equals to $\frac{2\pi}{\mid {\bf Q_f}\mid}$.

To study the surface state, we explore the Hamiltonian using the periodic boundary condition along the $x$ and $z$ directions, and the open boundary condition along the $y$-direction. The Hamiltonian can be rewritten as,
\begin{eqnarray}
H_W=&-\sum\limits_{{\bf k}{i_y}} [C^{\dagger}_{i_y}({\bf k}) t_y\sigma_3 C_{i_y+\hat{y}}({\bf k})+h.c.] \nonumber\\
&+\sum\limits_{{\bf k}{i_y}}\sum\limits_{\alpha={\hat{x},\hat{z}}}C^\dagger_{i_y}({\bf k})( h\sigma_3-2 t_\alpha\cos k_\alpha \sigma_3-\mu\sigma_0)C_{i_y}({\bf k})\nonumber\\
&+\sum\limits_{{\bf k}{i_y}}\lambda[C^\dagger_{i_y}({\bf k})2\sin k_x\sigma_1 C_{i_y}({\bf k})\nonumber\\&+C^\dagger_{i_y}({\bf k})\sigma_2 C_{i_y+\hat{y}}({\bf k})+h.c.].
\end{eqnarray}
Here the vector ${\bf k}$ represents a site in the reduced two-dimensional momentum space with ${\bf k}=(k_x,k_z)$.

The superconducting pairing part is rewritten as,
\begin{equation}
H^{s}_{sc}=\sum_{i_y{\bf k}}[\Delta_{0} c^\dagger_{i_y\uparrow}({\bf k})c^\dagger_{i_y\downarrow}(-{\bf k})+h.c.];
\end{equation}
\begin{eqnarray}
H^{p+ip}_{SC}=&\sum_{i_y{\bf k}\sigma}[\Delta_0\sin k_xc^{\dagger}_{{i_y}\sigma}({\bf k})c^{\dagger}_{{i_y}\sigma}({-\bf k})+h.c.\nonumber\\
&+\frac{\Delta_0}{2}c^{\dagger}_{{i_y}\sigma}({\bf k})c^{\dagger}_{{i_y+\hat{y}}\sigma}({-\bf k})+h.c.\nonumber\\&-\frac{\Delta_0}{2}c^{\dagger}_{{i_y}\sigma}({\bf k})c^{\dagger}_{{i_y-\hat{y}}\sigma}({-\bf k})+h.c.];
\end{eqnarray}
\begin{eqnarray}
H^{FFLO}_{SC}=&\sum_{i_y{\bf k}}[\Delta_0c^{\dagger}_{{i_y}\uparrow}({\bf k})c^{\dagger}_{{i_y}\downarrow}(-{\bf k}+{\bf Q_f})+h.c.\nonumber\\
&+\Delta_0c^{\dagger}_{{i_y}\uparrow}({\bf k})c^{\dagger}_{{i_y}\downarrow}(-{\bf k}-{\bf Q_f}).]+h.c.
\end{eqnarray}

The whole Hamiltonian can be expressed as the $4N_y\times 4 N_y$ (or $4rN_y\times 4r N_y$ for the FFLO state) matrix form.
One can obtain the spectral functions depending on the reduced momentum ${\bf k}$ and $y$, expressed as,
\begin{equation}
A_{{y}}({\bf k},\omega)=\sum_{\eta,\sigma} \frac{\mid u^{\eta}_{{i_y}\sigma}({\bf k})\mid^2}{\omega-E_\eta({\bf k})+i\Gamma},
\end{equation}
where $u^{\eta}_{{i_y}\sigma}$ and $E_\eta({\bf k})$ are eigenvectors and eigenvalues which can be obtained by diagonalizing the Hamiltonian matrix. Then the $y$ dependent LDOS is expressed as,
\begin{equation}
\rho_{y} (\omega)=\sum_{\bf k}A_{y}({\bf k},\omega).
\end{equation}

We are able to look into further the bulk state via the Fourier transformation, such that the Hamiltonian $H_W$ is given by
\begin{eqnarray}
H_{W}=&\sum\limits_{{\bf k},\alpha\sigma}(-2\sigma t_\alpha\cos k_\alpha-\sigma h-\mu)c^\dagger_{{\bf k}\sigma}c_{{\bf k}\sigma}\nonumber\\
&+ \sum_{\bf k}2\lambda [(\sin k_x + i\sin k_y)c^\dagger_{{\bf k}\uparrow}c_{{\bf k}\downarrow}+h.c.],
\end{eqnarray}
where ${\bf k}=(k_x,k_y,k_z)$ represents a site in the three dimensional momentum space. Correspondingly, the superconducting part can be written as
\begin{equation}
H^{s}_{sc}=\sum_{\bf k}(\Delta_0 c^\dagger_{{\bf k}\uparrow}c^\dagger_{-{\bf k}\downarrow}+h.c.),
\end{equation}
or
\begin{equation}
H^{p}_{sc}=\sum_{{\bf k}\sigma}[2\Delta_0 (\sin k_x+i\sin k_y) c^\dagger_{{\bf k}\sigma}c^\dagger_{-{\bf k}\sigma}+h.c.],
\end{equation}
or
\begin{equation}
H^{FFLO}_{sc}=\sum_{\bf k}(\Delta_0 c^{\dagger}_{{\bf k}\uparrow}c^{\dagger}_{{\bf -k}+{\bf Q_f}\downarrow}+\Delta_0 c^{\dagger}_{{\bf k}\uparrow}c^{\dagger}_{{\bf -k}-{\bf Q_f}\downarrow}+h.c.).
\end{equation}

The above Hamiltonian can be expressed as the $4\times 4$ (or $4r\times 4r$) matrix.
The Green's function is defined by diagonalizing the Hamiltonian, with its elements being expressed as,
\begin{equation}
\hat G_0 ({\bf{k}},\omega )_{ij}=\sum\limits_n {\frac{{u_{i\eta} ({\bf{k}})u_{\eta j}^\dag  ({\bf{k}})}}{{\omega  - E_\eta ({\bf k})  + i\delta }}}.
\end{equation}
$u_{ij} ({\bf{k}})$ and $E_\eta ({\bf k})$ are eigen-wave-vectors and eigen-values of the Hamiltonian matrix, respectively.

We now consider a single impurity placed on the site $(0,0,0)$ with the impurity scattering strength $V_s$.  The $T$-matrix is defined as,
\begin{equation}
\hat T(\omega ) = {{\hat U_0 } \mathord{\left/
 {\vphantom {{\hat U_0 } {\left[ {\hat I_{4(4r)} - \hat U_0 \frac{1}{N}\sum\limits_{\bf{k}} {\hat G_0 ({\bf{k}},\omega )} } \right]}}} \right.
 \kern-\nulldelimiterspace} {\left[ {\hat I_{4\times4(4r\times 4r)} - \hat U_0 \frac{1}{N}\sum\limits_{\bf{k}} {\hat G_0 ({\bf{k}},\omega )} } \right]}},
 \end{equation}
  with $\hat{I}$ is the identity matrix and $\hat U_0=V_s \hat\sigma_3 \otimes \hat I$.

The LDOS near the impurity is then expressed as,
\begin{equation}
\rho ({\bf r},{\omega})  =  - \frac{1}{\pi }{\mathop{\rm Im}\nolimits} {\rm{Tr}}\hat G ({\bf{r}},\omega ),
\end{equation}
with
\begin{equation}
\hat G({\bf{r}},\omega ) = \hat G_0 (0,\omega ) + \hat G_0 ({\bf{r}},\omega )\hat T(\omega )\hat G_0 ( - {\bf{r}},\omega ).
\end{equation}
The bare Green's functions $\hat G_0({\bf{r}},\omega )$ in the real space can be obtained by performing a Fourier transformation to [$\hat G_0 ({\bf{k}},\omega )$].

For the results to be presented below,
the input parameters are chosen as: $t_x=t_y=0.5$, $t_z=1$, $\Delta_0=0.2$, $\lambda=0.5$, $\mu=0.5$, and $h=2+2\cos(\pi/4)$~\cite{tzhou}.
With these parameters, there are
two Weyl points at $(0,0,\pm\pi/4)$.
In the FFLO pairing state, the Cooper pair momentums are $(0,0,\pm \pi/2)$.
Setting the lattice constant as the length unit, the real space period $r$ of the FFLO state equals to four.

\section{Results and discussion}

\subsection{The spectral function}

 \begin{figure}
\centering
  \includegraphics[width=2.6in]{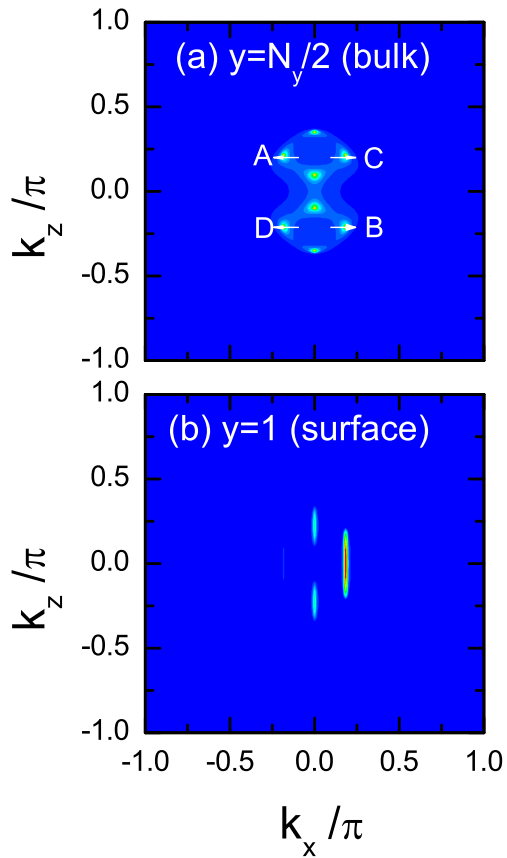}
\caption{(Color online) Intensity plots of the zero energy spectral functions for the $p+ip$-wave pairing symmetry. The open boundary along the $y$-direction for $N_y=100$ is considered.
}
\end{figure}

The spectral function may provide direct information for the quasiparticle energy bands. Especially, the spectral weight of each energy band can be obtained through investigating the spectral function. For multiband system the spectral weight of different energy bands is important.
And for our present work, due to the existence of the spin flip hopping when $\lambda\neq 0$, our starting model is an effective multi-band model thus the studies of the spectral function are indeed necessary. On the other hand, for the FFLO state the effective Brillouin zone is reduced and the bands are folded. The spectral weight of each band is particularly important to explore the electronic structure.

 \begin{figure}
\centering
  \includegraphics[width=3in]{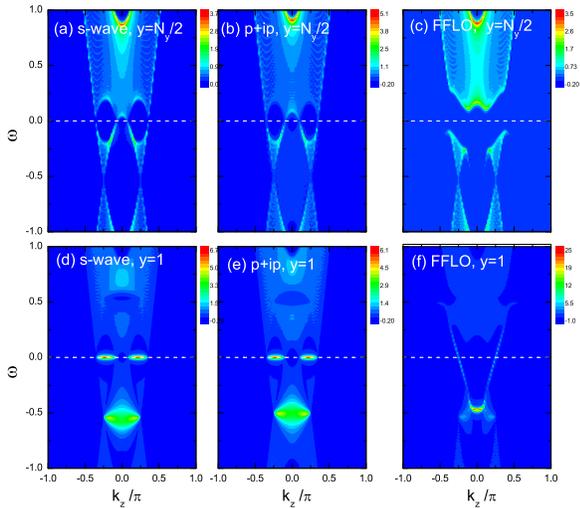}
\caption{(Color online) Intensity plots of the spectral function as a function of the $z$-direction momentum $(k_z)$ and the energy $\omega$ for $k_x=0$.
}
\end{figure}

We first discuss the numerical results for the spectral function at the Fermi energy ($\omega=0$).
For the BCS-like $s$-wave pairing symmetry and the FFLO pairing state, this issue has been addressed before~\cite{meng,cho,tzhou}. For the bulk spectra of the $s$-wave pairing symmetry, there are four nodal points along the $k_z$-axis.
Here we  focus only on the case of the lightly doped and weak superconducting pairing strength. When the superconducting order parameter or chemical potential increases, the bulk nodes may reduce and disappear completely for a rather strong superconducting pairing~\cite{meng,tzhou}, while this issue is not concerned with in the present work.
 At the system surface, three connected Fermi arcs exist~\cite{tzhou}. In the FFLO-state, the system in the bulk is fully gapped. At the system surface, there is one Fermi arc connecting the two Weyl points. The differences of the bulk states between the $s$-wave state and the FFLO-state may well be understood based on the spin texture picture~\cite{cho}.

We now present the numerical results for the $p+ip$-wave pairing symmetry.
For the open boundary condition along the $y$-direction and the periodic boundary condition along the $x$ and $z$ directions, the intensity plots of the zero energy $y$-dependent spectral functions are illustrated in Fig.~1. In the system bulk, as is seen in Fig.~1(a), there are eight bulk nodes in the Brillouin zone. The origin of the nodes can well be  understood from the gap formula of the $p+ip$-wave and the spin texture. The spin states for points $A$ and $B$ ($C$ and $D $) are antiparallel, as indicated by the arrows. For the case of $p+ip$-wave with the equal-spin pairing, only electrons with parallel spins can be paired.
Thus the four nodal points ($A$, $B$, $C$, and $D$) are left. On the other hand, the superconducting gap equals to zero when both $k_x$ and $k_y$ equal to zero, which generate the other four nodes along the axis of $k_x=0$.
At the system surface, as is seen in Fig.~2(b), there are three disconnected Fermi arcs, among which two are along the $k_z$-axis  and one is along the $k_x=0.2\pi$ line. This result is different from the the spectra of the $s$-wave state and FFLO state~\cite{tzhou}.
These features may be detected by later ARPES experiments and used to distinguish different pairing states of Weyl superconductors.

 \begin{figure}
\centering
  \includegraphics[width=3in]{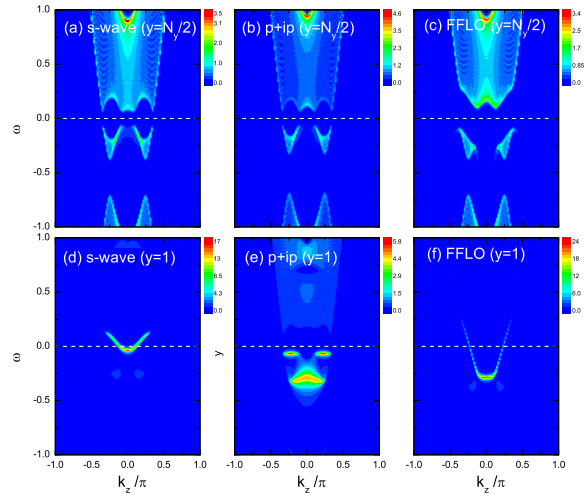}
\caption{(Color online) Similar to Fig.~2 but for $k_x=0.06\pi$.
}
\end{figure}

We here look into the spectral function at finite energies.
The intensity plots of the spectral function as a function of the momentum $k_z$ and the energy with $k_x=0$
 are displayed in Fig.~2. The spectral functions in the system bulk $(y=N_y/2)$ for different superconducting pairing states are seen in Figs.~2(a)-2(c). Generally, the spectra are qualitatively similar when the energy is far away from the Fermi energy, while they are rather different at low energies. An obvious energy gap (about 0.1) opens for the FFLO state. At the system surface $(y=1)$, as is seen in Figs.~2(d)-2(f),
the spectral weight transfers to the surface states and the bulk states are nearly invisible. For the case of the BCS-type pairing, only three flat surface bands exist, among which two are at the Fermi energy, connecting the bulk nodes. The other band is near the chemical potential energy, due to the normal state arc state at $\omega=\mu$. For the FFLO state, the surface states cross the Fermi energy, connecting the upper and the lower bulk bands. This is significantly different from those in the BCS-type states. The surface states shown in Fig.~2(f) can be understood well through analysing its topological feature, namely, it can be characterized by the
integer-valued three dimensional
topological superconductivity, which accounts for the existence of the surface states~\cite{yli}. The BCS-type pairing states and the FFLO state can be resolved
clearly from the above features. While in the BCS states, the spectral functions for different pairing symmetries ($s$-wave and $p+ip$-wave) are almost the same, thus they cannot be distinguished merely from the spectral functions along $k_x=0$ direction.

To resolve different pairing states further, we study the spectral function along a different line cut.
The intensity plots of the spectral function as a function of the momentum $k_z$ and the energy $\omega$ with $k_x=0.06\pi$ are explored in Fig.~3.
As is seen, for the bulk states, there exist energy gaps at the Fermi energy and the chemical potential energy for all of the three superconducting pairing states. At the system surface,
for the $p+ip$ pairing symmetry, there still exist several flat arc states, while their energies shift compared with those along the $k_x=0$ line.
For the FFLO state,
the surface state is qualitatively similar to that along $k_x=0$ line. However, the surface state for the $s$-wave pairing symmetry is completely different, namely, the flat arc states disappear and there are surface states crossing the Fermi energy and connecting two disconnected bulk bands
As a result, the surface spectra shown in Figs.~3(d)-3(f) are significantly different for the three kinds of superconducting state, which may be used to differentiate further the possible pairing states of the Weyl superconductor.

\subsection{Local density of states}

At this stage, we turn to address the LDOS spectra. The LDOS spectra can be detected experimentally through STM experiments, which is also a powerful tool to study the pairing symmetries of unconventional superconducting.   It was also proposed before that the LDOS spectral might be used to confirm the FFLO state through the periodic intensities in the real space or the in-gap Andreev bound states~\cite{qwang,tzhou2}.

 \begin{figure}
\centering
  \includegraphics[width=3in]{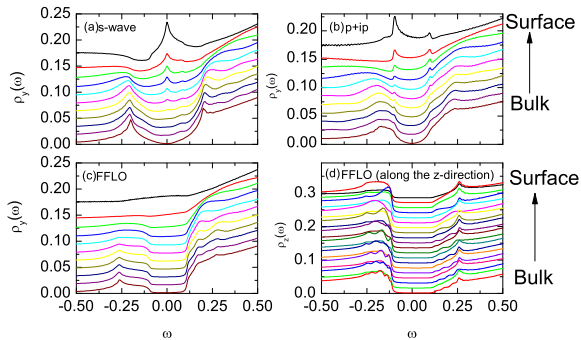}
\caption{(Color online) The LDOS spectra from the system bulk to the surface for different superconducting pairing states.
}
\end{figure}

For the open boundary along the $y$-direction and periodic boundary along the $x$ and $z$ directions, the $y$-dependent LDOS spectra $\rho_y(\omega)$ for different superconducting pairing states are displayed
in Figs.~4(a)-4(c). We first look into the bulk state. As is seen, for the BCS-type pairing states, the spectra are $V$-shaped, indicating the nodal behavior. For the FFLO state, $U$-shaped spectra are displayed with the intensity being nearly zero at a low energy region, indicating the fully gap feature.  Notably, the spectra for different pairing states are very different near the system surface. For the $s$-wave state, a clear zero energy peak shows up. For the $p+ip$-wave symmetry, a gap-like feature exists at low energies. For the FFLO state, the spectra is smooth across the Fermi energy. No peak and gap features exist. Thus the surface states along the $y$-direction may be used to distinguish different superconducting states.

For the FFLO state with the net Cooper pair momentum along the $k_z$ direction, the superconducting order parameter is periodic along the $z$-axis in the real space. It is usually useful to investigate the LDOS spectral along this direction to look for additional signatures for the FFLO states~\cite{qwang,tzhou2}. As for the open boundary along $z$-direction, we here present the numerical results of $z$-dependent the LDOS spectra $\rho_z(\omega)$ in Fig.~4(d).
It seems that the LDOS spectra depend weakly on $z$. From the bulk to the surface, there is no clear periodic feature for the LDOS. And no in-gap Andreev bound states appear. These features are different from the conventional FFLO state~\cite{qwang,tzhou2}. Thus for the Weyl superconductors, the FFLO state cannot be detected directly through its periodic feature or in-gap Andreev bound states.

 \begin{figure}
\centering
  \includegraphics[width=3in]{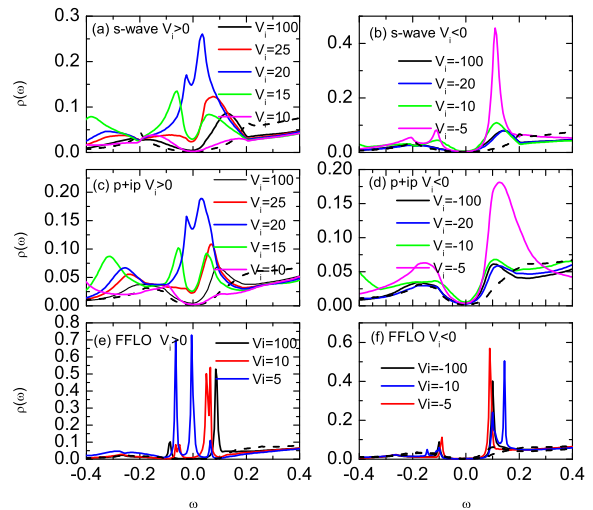}
\caption{(Color online) The LDOS spectral near a point impurity for different pairing states and different impurity scattering strengths.
}
\end{figure}

Now let us study the single impurity effect for the Weyl superconductor.
Usually for the topological nontrivial systems, at the system surface no impurity resonance states can survive due to the existence of the surface states.
While for the Weyl system, if the the cleavage surface is perpendicular to the $z$-axis,
 there is no surface states and
 the LDOS spectra are qualitatively the same with the bulk ones. For this case the
single impurity effect can be explored based on Eqs.~(17-19).
The LDOS spectra for different superconducting pairing states with different impurity scattering strength are presented in Fig.~5. For the BCS-type pairing and positive impurity scattering, strong resonance peaks appear near the Fermi level for a typical impurity strength $V_i=20$. The intensity of the impurity induced peaks decreases rapidly when the impurity strength is away from $20$ and finally the impurity states nearly disappear. For the negative scattering potentials, the strong resonance states appear near the impurity strength $V_s=-10$ while their positions are near the gap edges. Similar to the case of positive potentials, the impurity induced peaks decreases rapidly when the strength away from $V_s=-10$. For the FFLO state, the impurity resonance states are rather robust.
The strong impurity induced peaks appear for nearly all of the impurity scattering strengths we considered.
For the BCS states with different pairing symmetries ($s$-wave and $p$-wave), no qualitative differences exist, as shown in Figs.~5(a-d). While the results are significantly different for the FFLO superconducting state.
Thus we here proposed that the impurity effect may also help to differentiate the BCS states from the FFLO state.
While it cannot resolve different pairing symmetries of the BCS states.


Generally, the impurity induced in-gap resonance peaks is phase sensitive and usually they may be robust for a strong impurity scattering strength. While here for the BCS-type pairing, it seems that the impurity resonance states just appear occasionally and are not rather
robust. Actually, the effective order parameter phase of a Weyl superconductor has been studied systematically recently~\cite{yli}. For the BCS
type pairing, the phase cannot be
globally well-defined on the Fermi surfaces. Thus the impurity effect displayed above in the BCS-type states is understandable. On the contrary, in the FFLO state,
the phase is well-defined. Especially, the effective pairing function in this pairing state is a mixture of the $s$-wave pairing and $p$-wave pairing. It can be characterized
by an integer topological index. The single impurity effect in various topological superconductors has been studied intensively, and the in-gap impurity states have been confirmed and well understood~\cite{jsau,xjliu,nagai,wimm,hhu,nag,guo}. Our numerical results about the impurity effect in the FFLO state is well consistent with previous theoretical results for topological superconductors.

\section{summary}

In summary, the single particle spectra of a Weyl superconductor are studied numerically. We have addressed three different superconducting pairing states,
i.e., the BCS-like pairing with the $s$-wave pairing symmetry, the BCS-like pairing with the $p+ip$-wave pairing symmetry, and the FFLO pairing state.
For the open boundary condition along the $y$-direction, the bulk states and surface states for the spectral function and the LDOS spectra are
explored, which can be used to distinguish the three supercnducting pairing states clearly.
Along the $z$-axis there are no surface states. The low energy spectra near the system surface is qualitatively the same with those in the system bulk.
The single impurity effect is investigated.
 For the BCS-type pairing state, the impurity states appear at certain typical impurity scattering strength and the intensities decrease rapidly when away from this strength.
 For the FFLO state the impurity induced in-gap peaks are strong and robust, and they appear for all of the impurity scattering strengths we considered. These results may be used to differentiate the BCS-type pairing states from the FFLO pairing state. It is expected that the present results will be rather helpful for a comprehensive and better understanding of Weyl superconductors.

This work was supported by the Start-up Foundation from South China Normal University and the Natural Science Foundation from Jiangsu Province of China (Grant No. BK20160094).


\end{document}